\documentclass{article}
\usepackage{graphics,url}
\usepackage{amsfonts, amsmath, amssymb, xcolor,setspace}
\usepackage[]{algorithm, algorithmic}
\usepackage{graphicx}
\usepackage{hyperref}
\usepackage{float}

\usepackage[margin=1.25in]{geometry}

\usepackage{listings}
\usepackage{color}
\definecolor{dkgreen}{rgb}{0,0.6,0}
\definecolor{gray}{rgb}{0.5,0.5,0.5}
\definecolor{mauve}{rgb}{0.58,0,0.82}
\newcommand{\argmin}{\mathop{\mathrm{argmin}}\limits}
\newcommand{\minF}{\mathop{\mathrm{min}}\limits}
\newcommand{\maxF}{\mathop{\mathrm{max}}\limits}
\newcommand{\aveF}{\mathop{\mathrm{ave}}\limits}

\newcommand{\bx}{\textbf{x}}

\newcommand{\bw}{\textbf{w}}
\newcommand{\bz}{\textbf{z}}
\newcommand{\bh}{\textbf{h}}
\newcommand{\btheta}{\mbox{\boldmath${\theta}$} }
\def\TT{{\mbox{\tiny T}}}

\usepackage[normalem]{ulem}

\usepackage{authblk}

\begin{document}
\title{Statistical Analysis of Complex Computer Models in Astronomy}
\author[1]{Joshua Lukemire}
\author[2]{Qian Xiao}
\author[2]{Abhyuday Mandal\thanks{{amandal@stat.uga.edu}}}
\author[3]{Weng Kee Wong}
\affil[1]{Department of Biostatistics and Bioinformatics, Emory University, Atlanta, GA, USA}
\affil[2]{Department of Statistics, University of Georgia, Athens, GA, USA}
\affil[3]{Department of Biostatistics, University of California, Los Angeles, CA, USA}
\date{}                     %% if you don't need date to appear

\setcounter{Maxaffil}{0}
\renewcommand\Affilfont{\itshape\small}
%  \inst{2}\fnmsep\thanks{\email{amandal@stat.uga.edu}}
% \footremember{eml}{\email{amandal@stat.uga.edu}}

%\doublespacing

\maketitle

\abstract{
We introduce statistical techniques required to handle complex computer models with potential applications to astronomy. Computer experiments play a critical role in almost all fields of scientific research and engineering. These computer experiments, or simulators, are often computationally expensive, leading to the use of emulators for rapidly approximating the outcome of the experiment. Gaussian process models, also known as Kriging, are the most common choice of emulator. While emulators offer significant improvements in computation over computer simulators, they require a selection of inputs along with the corresponding outputs of the computer experiment to function well. Thus, it is important to select inputs judiciously for the full computer simulation to construct an accurate emulator. Space-filling designs are efficient when the general response surface of the outcome is unknown, and thus they are a popular choice when selecting simulator inputs for building an emulator. In this tutorial we discuss how to construct these space filling designs, perform the subsequent fitting of the Gaussian process surrogates, and briefly indicate their potential applications to astronomy research. 
}

%%%%%%%%%%%%%%%%%%%%%%%%%%%%%%%%%%%%%%%%%%%%%%%%%%%%%%%%%%%%%%%%%%%%%%%%%%%%%%%%%%%%%%%%%
%                                                                                       %
%                                                                                       %
%                                                                                       %
%                                     SECTION 1                                         %
%                                                                                       %
%                                                                                       %
%                                                                                       %
%%%%%%%%%%%%%%%%%%%%%%%%%%%%%%%%%%%%%%%%%%%%%%%%%%%%%%%%%%%%%%%%%%%%%%%%%%%%%%%%%%%%%%%%%
\section{Introduction}\label{intro}

Computer experiments, or simulators, are an increasingly important tool in many scientific fields. In these experiments, a computer model is defined relating a set of inputs to an output. Instead of conducting a traditional experiment, a researcher will provide a set of inputs to the computer model and obtain the model output. This approach is very appealing in fields such as physics, where the computer experiment model can be setup using a series of  known relationships/equations and different inputs may consist of unknown constants in those equations or other properties such as mass or chemical compositions. These experiments can be effective alternatives to experiments which may be too expensive or otherwise impossible to perform in a traditional setting. They differ from standard experiments in several key ways. Most importantly, computer experiments are generally deterministic; for a set of input settings the experiment will return the same result every time it is conducted. Second, the experiments will generally not have an easily described response surface; for example a standard linear regression model will not generally describe the outcome accurately. 

\begin{figure}[h]
    \centering
    \includegraphics[width=0.35\linewidth]{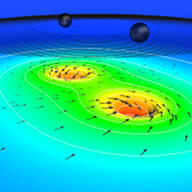}
    \caption{An example from a simulation examining whether two black holes merge. Source: \url{https://www.black-holes.org/code/SpEC.html}.}
    \label{fig:my_label}
\end{figure}

Many research areas in astronomy do not easily permit conducting traditional experiments. For example, researchers may be interested in the formation of binary black holes. Clearly the researchers will not be able to create multiple black holes and observe their dynamics over time. Computer experiments make it possible to study such phenomena by creating computer models based on the theorized properties of these binary systems and then comparing the output to what is observed in Nature. For example, Compact Object Mergers: Population Astrophysics and Statistics (COMPAS) is used to investigate binary population synthesis. The computer experiment takes input as initial conditions and simulates the lifespan of stars \cite{Stevenson1} \cite{COMPAS2}. Similarly, binary population synthesis code ComBinE has been used to perform binary population syntheses \cite{kruckow2018progenitors}, and the tool UniverseMachine \cite{behroozi2019universemachine} allows researchers to study galaxy formation.

% DO NOTE DELETE; COMPAS REFERENCES: \bigskip\noindent Relevant references: \ref{Stevenson1, COMPAS2}

% DO NOT DELETE - SPECTRAL EINSTEIN LINK\bigskip\noindent Spectral Einstein Code. fully general-relativistic compact object simulations. \url{https://www.black-holes.org/code/SpEC.html}. NOTE THIS IS NOT COMPUTER EXPERIMENT ITSELF, BUT IS USED IN COMPUTER EXPERIMENTS

Computer experiments for many complex systems  can be very expensive to perform (see, for example, \cite{williams2019}).  This computational expense can be a significant problem, especially if a researcher hopes to conduct the experiment for many sets of inputs. An alternative to directly performing these computer experiments is to instead create a surrogate or emulator \cite{Gramacy2020}. Surrogates are popular for computer experiments when it is not realistic to evaluate a fine grid over the entire input space. Instead, a (relatively) small number of points are chosen to evaluate under the original computer simulation. Then, a model is fit to the output from these limited runs. Predictions under this model for new inputs, as well as  uncertainty quantification, can be obtained from the surrogate without the need to re-run the expensive computer simulation at the new points. If the model fits well, then the predicted value will be close to the true value that would have been obtained if the full computer experiment was used. 

The most common tool used to fit the data points and create the surrogate model is the Gaussian process (GP)  \cite{Sacks89} \cite{Gramacy2020}. The GP is appealing for creating surrogates because it interpolates known data to evaluate new data points. This is especially important when the outcome for a fixed set of inputs is deterministic, which is frequently the case in computer experiments. This approach is becoming more popular in the astronomy literature. Some recent work includes \cite{gration2019dynamical}, who proposed using Gaussian process emulation to obtain confidence intervals for the parameter vector of a phase-space distribution function for dwarf spheroidal galaxies. 

Section 2 of this tutorial paper introduces Gaussian process models and discusses their applications to computer experiments. We provide codes and examples throughout in the R programming language \cite{RSoftware}. Section 3 of this paper focuses on determining what inputs to use to generate the responses used to fit the Gaussian process model to obtain an accurate surrogate. We draw upon the design of experiments statistical literature to discuss design of computer experiments. In particular we focus on Latin hypercube designs and discuss several techniques for finding them.

%%%%%%%%%%%%%%%%%%%%%%%%%%%%%%%%%%%%%%%%%%%%%%%%%%%%%%%%%%%%%%%%%%%%%%%%%%%%%%%%%%%%%%%%%
%                                                                                       %
%                                                                                       %
%                                                                                       %
%                                     SECTION 2                                         %
%                                                                                       %
%                                                                                       %
%                                                                                       %
%%%%%%%%%%%%%%%%%%%%%%%%%%%%%%%%%%%%%%%%%%%%%%%%%%%%%%%%%%%%%%%%%%%%%%%%%%%%%%%%%%%%%%%%%
\section{Surrogates for Computer Models}

\begin{figure}
    \centering
    \includegraphics[width=0.45\textwidth]{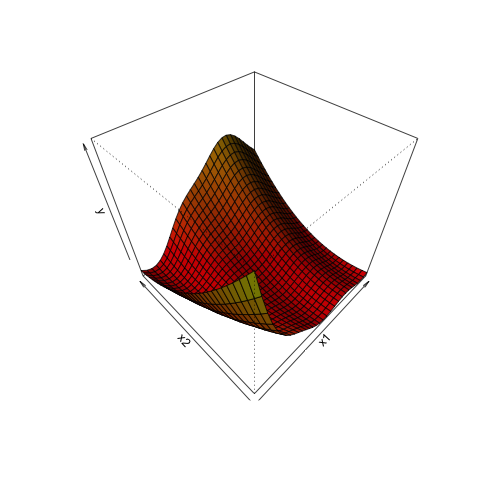}
    \includegraphics[width=0.45\textwidth]{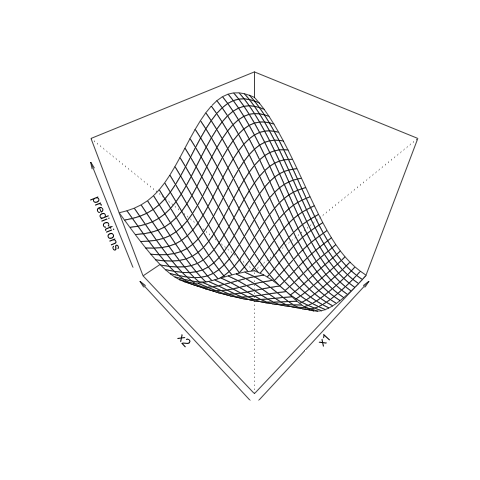}
    \caption{Left: the true response under the Branin function. Right: The estimated response using the surrogate model.}
    \label{fig:branintrue}
\end{figure}

\medskip\noindent Simpler surrogates or {\it emulators} are often preferred for complex deterministic computer models. Gaussian Process (GP) models are popular choice for this purpose \cite{Sacks89}. Consider an $n$-run computer experiment with $d$-dimensional input vectors $\bx_i = (x_{i1},\ldots,x_{id})^\TT$ and a deterministic output $y(\bx_i)$, for $i=1,2,\ldots,n$. To fix ideas, assume that we are interested in a 2-dimensional input for a computer experiment with output given by the Branin function as defined by \cite{Derek2}, see also \cite{forrester2008engineering}.

\begin{equation}
    y(x_1, x_2) = \left(x_2 - \frac{5.1}{4 \pi^2} x_1^2 + \frac{5}{\pi} x_1 - 6 \right)^2 + 10 \left(1 - \frac{1}{8 \pi} \right)\cos \left( x_1 \right) + 10,
\end{equation}
where the design space is given by values of $x_1 \in [-5, 10]$ and $x_2 \in [0, 15]$. The R code below can be used to evaluate this function.
\begin{lstlisting}
# x is a vector of inputs (length 2)
branin <- function(x){
  a <- 1
  b <- 5.1  / (4 * pi^2)
  c <- 5 / pi
  r <- 6
  s <- 10
  t <- 1 / (8 * pi)
  return( a*(x[2] - b*x[1]^2 + c*x[1] -r)^2 + s*(1-t)*cos(x[1]) + s )
}
\end{lstlisting}
The left panel in Figure \ref{fig:branintrue} displays the output for this function over the entire design space.

%%%%%%%%%%%%%%%%%%%%%%%%%%%%%%%%%%%%%%%%%%%%%%%%%%%%%%%%%%%%%%%%%%%%%%%%%%%%%%%%%%%%%%%%%
%                                                                                       %
%                                                                                       %
%%%%%%%%%%%%%%%%%%%%%%%%%%%%%%%%%%%%%%%%%%%%%%%%%%%%%%%%%%%%%%%%%%%%%%%%%%%%%%%%%%%%%%%%%
\subsection{Stationary Gaussian Process - Krigging}

The simplest possible GP model, known as ordinary GP or $krigging$, is given by
\begin{equation}\label{eq:GP}
    y(\bx_i) = \mu + Z(\bx_i),
\end{equation} 
where $\mu$ is the mean and $Z(\bx)$ is a GP, denoted by $Z(\bx)\sim GP(0, \sigma^2R)$. This notation implies that the GP has zero-mean, and the covariance function $Cov\left(Z(\bx_i), Z(\bx_j)\right) =  \sigma^2 R(\cdot|\btheta)$, where $\btheta=(\theta_1,\ldots,\theta_d)^\TT$ is the vector of unknown correlation parameters with all $\theta_s > 0$ ($s =1, \ldots, d$). The correlation between outputs is determined by a stationary correlation function $R$ with parameter $\btheta$. Two of the more commonly-used correlation functions are the power-exponential and the Gaussian functions. Under a power-exponential correlation structure the $(i,j)^{\mbox{th}}$ term is defined as:
\begin{eqnarray}\label{eq:1}
R(\bx_{i},\bx_{j}|\btheta) =  \prod_{s = 1}^{d} \exp\Bigg\{-\theta_s \mid x_{is} - x_{js} \mid ^{p_s} \Bigg\} \hspace{0.2in} \mbox{for all }  i,j,
\end{eqnarray}
where smoothness parameters $p_1, \ldots, p_s$ are all between $0$ and $2$. Of special importance is $p_s=2$, for all $s = 1, \ldots, d$, which corresponds to the popular Gaussian correlation function:
\begin{equation}
\label{gc}
R(\bx_{i},\bx_{j}|\btheta) = \hbox{exp} \left\{ - \sum_{s=1}^d \theta_s  (x_{is} - x_{js})^2\right\} \hspace{0.2in} \mbox{for all }  i,j.
\end{equation}

The flexibility of the correlation structure is what makes the GP model a popular surrogate for complex computer models. For any given input $\bx^*$ in the design space, the fitted GP surrogate gives the predicted computer model response as,
\begin{equation}
	\label{eqn-gp-estimation}
	\hat{y}(\bx^*) = \mu + \mathbf{r}^\TT(\bx^*)\mathbf{R}^{-1}(\mathbf{y}-\mu\mathbf{1}_n),
\end{equation}
where 
\begin{equation}
\mathbf{r}(\bx^*) = \Bigg[\text{corr}\Big(Z(\bx^*), Z(\bx_1)\Big), \text{corr}\Big(Z(\bx^*), Z(\bx_2)\Big),\ldots,\text{corr}\Big(Z(\bx^*), Z(\bx_n)\Big)\Bigg]^\TT,
\end{equation}
$\mathbf{1}_n$ is a vector of ones of length $n$, $\mathbf{R}$ is the $n\times n$ correlation matrix for $(Z(\bx_1), ..., Z(\bx_n))^\TT$,  $\mathbf{y}$ is the response vector $(y(\bx_1),\ldots,y(\bx_n))^\TT$, and the associated uncertainty estimate is
\begin{equation}
	\label{eqn-gp-estimated-variance}
	s^2(\bx^*) = \sigma^2\Big(1 - \mathbf{r}(\bx^*)^\TT\mathbf{R}^{-1}\mathbf{r}(\bx^*)\Big).
\end{equation}
In practice, the parameters $\mu, \sigma^2$ and $\btheta$ in Equations~(\ref{eqn-gp-estimation}) and (\ref{eqn-gp-estimated-variance}) are unknown and need to be estimated from the data. The parameters can be estimated using the \texttt{mlegp} function in R. Assume that we already have a design with 10 points (details for obtaining this design will be presented in Section 3). Then we can fit a GP with a Gaussian correlation function as,

\begin{lstlisting}
library(mlegp)
# Obtaining this design is discussed in Section 3
design <- matrix(c(-4.25, -1.25, 4.75, 7.75, -2.75, 1.75, 
    6.25, 3.25, 0.25, 9.25, 8.25, 6.75, 11.25, 12.75, 0.75, 
    2.25, 9.75, 5.25, 14.25, 3.75), ncol=2)
# Obtain the output at the set points in our design
Yx <- apply(design, 1, branin)
# Use the observed outputs to construct a surrogate
branin_surrogate_1 <- mlegp(design, Yx)
\end{lstlisting}

Similarly, estimates across the entire design space can be obtained by using the surrogate model by specifying the inputs on a grid:

\begin{lstlisting}
# Construct a grid of points to obtain predictions at
x1 <- seq(from = -5, to =10, length.out = 25)
x2 <- seq(from = 0, to =15, length.out = 25)
test_points <- expand.grid(x1, x2)
# Get predictions from the gaussian process
yhat <- predict(branin_surrogate_1, test_points)
predictions <- matrix( yhat, nrow = length(x1) )
# plot the predictions, theta and phi control viewing angle
persp(x1, x2, predictions, theta = -45, phi=45)
\end{lstlisting}
The right panel in Figure \ref{fig:branintrue} displays a plot of the surrogate output. Comparing this output to the true values in the left panel, it is clear that the surrogate model is able to obtain a very close approximation to the true process.

\bigskip\noindent The formulation in equation~(\ref{eq:GP}) can be extended to incorporate a global trend function for the mean $\mu$ \cite{Wackernagel}. This is known as {\it Universal Kriging}:
\begin{equation}\label{eq:GPu}
y(\bx) = \mu(\bx) + Z(\bx),
\end{equation} 
with $\mu(\bx) = g(\bx)^\TT\beta = \sum_{i=1}^m \beta_i g_i(\bx)$, where $g$ is a $m$-dimensional known function and $\beta=(\beta_1,\ldots,\beta_m)^\TT$ is a vector of unknown parameters. If we assume $g_1(\bx)=1$ and let $\mathbf{G} = (g_1(\bx),\ldots,g_m(\bx))^\TT$, then the optimal predictor under model~(\ref{eq:GPu}) is given by
\begin{equation}\label{ugp-estimation}
	\hat{y}(\bx^*) = g^\TT(\bx^*)\hat{\beta} + \mathbf{r}^\TT(\bx^*)\mathbf{R}^{-1}(\mathbf{y}-\mathbf{G}\hat{\beta}),
\end{equation}
where $\hat{\beta}= (\mathbf{G}^\TT \mathbf{R}^{-1} \mathbf{G})^{-1}(\mathbf{G}^\TT \mathbf{R}^{-1} \mathbf{y})$. If the assumed $\mu(\bx)$ is close to the truth, this formulation will lead to a better prediction than ordinary krigging. Note that this universal kriging formulation uses $\mu(\bx)$ to capture the known trends, but in most real applications, these trends are not known, and hence ordinary kriging is commonly used \cite{Welch1992}.

%%%%%%%%%%%%%%%%%%%%%%%%%%%%%%%%%%%%%%%%%%%%%%%%%%%%%%%%%%%%%%%%%%%%%%%%%%%%%%%%%%%%%%%%%
%                                                                                       %
%                                                                                       %
%%%%%%%%%%%%%%%%%%%%%%%%%%%%%%%%%%%%%%%%%%%%%%%%%%%%%%%%%%%%%%%%%%%%%%%%%%%%%%%%%%%%%%%%%
\subsection{Non-stationarity}
Note that Equation~(\ref{eq:1}) refers to a stationary GP, that is 
\begin{equation}\label{eq:st}
Cov\Big(Z(\bx+\bh),Z(\bx)\Big) = \sigma^2 R(\bh),
\end{equation}
where the correlation function $R(\bh)$ is a positive semidefinite function with $R(\mathbf{0})=1$ and $R(-\bh)=R(\bh)$. These stationary Gaussian processes are popular surrogates for complex computer models, since it can be shown that the corresponding predictor of $\mu$ in equation~(\ref{eq:GP})
\begin{equation}\label{Eq:mu-hat}
\hat{\mu} = (\mathbf{1}_n^\TT \mathbf{R}^{-1} \mathbf{1}_n)^{-1} \mathbf{1}_n^\TT \mathbf{R}^{-1}\mathbf{y}
\end{equation}
is the best linear unbiased predictor (BLUP) in the sense that it minimizes the mean squared prediction error. In reality this assumption of stationarity may not hold. Under these circumstances, the above predictor is no longer optimal. Some literature is available to deal with non-stationary Gaussian processes for emulating computationally expensive functions. For example, \cite{xiong2007} introduced the idea of nonlinear mapping based on a parameterized density function, and \cite{Gramacy2008} proposed a Bayesian tree structure by dividing the design space into subregions. 

\cite{Ba2012} used composite Gaussian process (CGP) models to address the nonstationarity problem. In their formulation, the model takes the following form:
\begin{eqnarray}\label{Eq:cgp}
\nonumber y(\mathbf{x}) &=& Z_{\text {global }}(\mathbf{x})+Z_{\text {local }}(\mathbf{x}), \\
Z_{\text {global }}(\mathbf{x}) &\sim& \operatorname{GP}\left(\mu, \tau^{2} R_1(\cdot)\right),\\
\nonumber Z_{\text {local }}(\mathbf{x}) &\sim& \mathrm{GP}\left(0, \sigma^{2} R_2(\cdot)\right).
\end{eqnarray}
Here $Z_{\text {global }}(\mathbf{x})$ and $Z_{\text {local }}(\mathbf{x})$ are two stationary GPs that are independent of each other. Just as the universal kriging generalizes the ordinary kriging by adding a trend function  $\mu(\mathbf{x})$, the composite Gaussian process model given in equation (\ref{Eq:cgp}) is a further extension which adds a more flexible global trend component. The model was extended to incorporate the non-constant variance assumption as follows:
\begin{eqnarray}\label{Eq:cgp2}
\nonumber y(\mathbf{x}) &=& Z_{\text {global }}(\mathbf{x})+\sigma(\bx)Z_{\text {local }}(\mathbf{x}), \\
Z_{\text {global }}(\mathbf{x}) &\sim& \operatorname{GP}\left(\mu, \tau^{2} R_1(\cdot)\right),\\
\nonumber Z_{\text {local }}(\mathbf{x}) &\sim& \mathrm{GP}\left(0, R_2(\cdot)\right).
\end{eqnarray}
The model can be further extended for noisy data by adding a third GP (with zero correlation) to the model (\ref{Eq:cgp2}).

%%%%%%%%%%%%%%%%%%%%%%%%%%%%%%%%%%%%%%%%%%%%%%%%%%%%%%%%%%%%%%%%%%%%%%%%%%%%%%%%%%%%%%%%%
%                                                                                       %
%                                                                                       %
%%%%%%%%%%%%%%%%%%%%%%%%%%%%%%%%%%%%%%%%%%%%%%%%%%%%%%%%%%%%%%%%%%%%%%%%%%%%%%%%%%%%%%%%%
\subsection{Numeric Considerations - Local GP}
Note that the prediction involves the inversion of the $n\times n$ correlation matrix $\mathbf{R}$, where $n$ is the number of data points (see equation~(\ref{eqn-gp-estimation}) or (\ref{Eq:mu-hat}), for example). This is a big hurdle in implementing GPs. To overcome this problem, \cite{Gramacy2015} introduced the idea of local Gaussian Process approximation for large computer models. They provided a family of local sequential design schemes that dynamically define the support points of a GP predictor based on a local subset of the data. Their approach is different from that of $k$-nearest neighbours. The basic idea is simple, under the standard choices of the covariance structures the correlation between points is dependent on the distance between those points, with data points far from $\bx^*$ having very little effect on its prediction. Hence it is not a good use of computational resources to invert the full covariance matrix, as the elements corresponding to ``far away" points will contribute little to the prediction of $y(\bx^*)$. An interested reader should refer to  \cite{Gramacy2015} for the formulas of the GP predictor based on a local subset of data. The end result is a global predictor that takes advantage of modern multicore parallel computing tools.

%%%%%%%%%%%%%%%%%%%%%%%%%%%%%%%%%%%%%%%%%%%%%%%%%%%%%%%%%%%%%%%%%%%%%%%%%%%%%%%%%%%%%%%%%
%                                                                                       %
%                                                                                       %
%%%%%%%%%%%%%%%%%%%%%%%%%%%%%%%%%%%%%%%%%%%%%%%%%%%%%%%%%%%%%%%%%%%%%%%%%%%%%%%%%%%%%%%%%
\subsection{Extension to Qualitative Inputs}

The conventional GP models consider quantitative predictor variables only, but many computer experiments may have both quantitative and qualitative inputs. In order to construct an emulator with qualitative factors, a naive approach would be to create distinct GP models for data collected at the different level combinations of the qualitative factors. Clearly this approach has many limitations, particularly when there are several qualitative factors. There are some more advanced techniques to deal with such cases. To fix ideas, for an $n$-run computer model, denote the $k^{th}$ ($k = 1, \ldots, n$) data input as $\bw_{k} = (\bx_{k}^\TT, \bz_{k}^\TT)^\TT$ where $\bx_k = (x_{k1},\ldots,x_{kp})^\TT \in \mathbb{R}^{p}$ is the quantitative part and $\bz_k = (z_{k1},\ldots,z_{kq})^\TT \in \mathbb{N}^{q}$ is the qualitative part (coded in levels) of the input. Note here that previously $\bx$ denoted the input, which was entirely continuous. However, now $\bw$ denotes the entire input, with $\bx$ referring to the continuous part.  For these kind of problems,  a popular GP based model was introduced by \cite{qian2008}, among many others \cite{han2009}, \cite{zhou2011}, \cite{swiler2014surrogate}, \cite{zhang2015computer} and \cite{zhang2018latent}.
Specifically, an ordinary GP model with a multiplicative covariance function is considered (for any two inputs $\bw_1$ and $\bw_2$):
\begin{equation}
\label{corm1}
\hbox{Cov}(Z(\bw_1), Z(\bw_2)) =  \sigma^2  \prod_{j=1}^{q}\tau^{(j)}_{z_{1j}z_{2j}} R(\bx_{1},\bx_{2}|\btheta) ,
\end{equation}
where the parameter $\tau^{(j)}_{z_{1j}z_{2j}}$ represents the correlation between two levels ($z_{1j}$ and $z_{2j}$) in the $j^{th}$ qualitative factor $z^{(j)}$, and $R(\bx_{1},\bx_{2}|\btheta)$ is given before in equation~(\ref{gc}). Different choices of $\tau^{(j)}_{z_{1j}z_{2j}}$ lead to different types of correlation functions.  For example, an exchangable correlation function is obtained when $\tau^{(j)}_{z_{1j}z_{2j}}$ is some constant between $0$ and $1$. Alternatively, an additive GP model was proposed in  \cite{deng2017additive}, which adopts the following covariance function:
\begin{equation}
\label{dc}
\hbox{Cov}(Z(\bw_1), Z(\bw_2)) = \sum_{j=1}^{q}\sigma_j^2\tau^{(j)}_{z_{1j}z_{2j}}R(\bx_1, \bx_2 \vert \btheta^{(j)}),
\end{equation}
where $\sigma_j^2$ and $\btheta^{(j)}$ ($j = 1, \ldots, q$) are the process variance and correlation parameters, respectively, corresponding to $z^{(j)}$.

The methods above do not have good physical interpretation of the correlation structures. Motivated by this, \cite{Qian1} proposed an EzGP method based on ANOVA (Analysis of Variance) ideas to jointly model the quantitative and qualitative inputs:
\begin{equation}\label{eq:additiveGP}
Y(\bw) = \mu + G_{\bz}(\bx),
\end{equation}
which implies that for any level combination of $\bz$, $Y(\bw)$ is a Gaussian process. In particular, they considered
\begin{equation}
	\label{eq:agp}
	G_{\bz}(\bx) = G_{0}(\bx)+G_{z^{(1)}}(\bx) + \cdots + G_{z^{(q)}}(\bx),
\end{equation}
where $G_0$ and $G_{z^{(h)}}$ ($h = 1, \ldots q$) are independent Gaussian processes with mean zero and some covariance functions. Here, $G_{0}$ is a standard GP taking only quantitative inputs $\bx$, which can be viewed as the base GP reflecting the intrinsic relation between $y$ and $\bx$. On the other hand, $G_{z^{(h)}}$'s can be viewed as an adjustment to the base GP by the impact of the qualitative factor $z^{(h)}$ ($h = 1, \ldots q$). This EzGP technique enjoys some nice theoretical properties and is able to flexibly address heterogeneity in computer models involving multiple  qualitative factors. \cite{Qian1} also developed two variants of the EzGP model to achieve computational efficiency for data with high dimensionality and large sizes.

%%%%%%%%%%%%%%%%%%%%%%%%%%%%%%%%%%%%%%%%%%%%%%%%%%%%%%%%%%%%%%%%%%%%%%%%%%%%%%%%%%%%%%%%%
%                                                                                       %
%                                                                                       %
%%%%%%%%%%%%%%%%%%%%%%%%%%%%%%%%%%%%%%%%%%%%%%%%%%%%%%%%%%%%%%%%%%%%%%%%%%%%%%%%%%%%%%%%%
\subsection{Calibration}

The notion of calibration and sensitivity analysis is important in the context of physical and computer experiments. Instead of observing the real physical process, $y^{Real}$, we are only able to observe a process $y^{Field}$ as:
\begin{equation}\label{eq:F}
y^{Field}(\bx) = y^{Real}(\bx) + \epsilon,
\end{equation} 
where $\epsilon$ is the usual normal error. This $y^{Real}$ is approximated by a computer model $y^{Model}$. Note that the computer model $y^{Model}$ not only has the input variables $\bx$, but also some unknown parameters $\theta$, called calibration parameters which are used to fine tune the model. Note that these calibration parameters can be, for example, the correlation parameters discussed above. The field data $y^{Field}$ is used mainly to learn more about the real phenomenon $y^{Real}$. \cite{KOH} proposed a Bayesian framework to address this as follows:
\begin{eqnarray}\label{eq:KOH}
y^{Real}(\bx) &=& y^{Model}(\bx,\theta) + b(\bx) \\
\nonumber y^{Field}(\bx) &=& y^{Model}(\bx,\theta) + b(\bx) + \epsilon,
\end{eqnarray}
where $b(\bx)$ is a functional discrepancy, called bias. \cite{KOH} used Bayesian methods to estimate the bias correction function and unknown calibration parameter $\theta$ under a GP prior. An alternative to this Bayesian approach is an iterative history matching algorithm such as the one proposed by \cite{vernon2010} for calibrating a galaxy formation model called GALFORM. This is actually a hands-on process, which intelligently eliminates the implausible points from the input (or parameter) space and returns a set of plausible candidates for the parameters $\theta$. Recently, \cite{natalia} used this algorithm for calibrating hydrological time-series models and \cite{joseph} further extended this method with a more systematic approach, in which they discretize the target response series on a few time points, and then iteratively apply the history matching algorithm with respect to the discretized targets.

%%%%%%%%%%%%%%%%%%%%%%%%%%%%%%%%%%%%%%%%%%%%%%%%%%%%%%%%%%%%%%%%%%%%%%%%%%%%%%%%%%%%%%%%%
%                                                                                       %
%                                                                                       %
%                                                                                       %
%                                     SECTION 3                                         %
%                                                                                       %
%                                                                                       %
%                                                                                       %
%%%%%%%%%%%%%%%%%%%%%%%%%%%%%%%%%%%%%%%%%%%%%%%%%%%%%%%%%%%%%%%%%%%%%%%%%%%%%%%%%%%%%%%%%
\section{Design of Computer Experiments}

The computer experiments under consideration have deterministic outputs, and thus replicates at a given set of input settings should be avoided, as they do not provide any further information about the response. Good designs for computer experiments are then designs that are ``space-filling" in some sense, which make it easier to fit accurate surrogate models. We will next discuss a few types of space-filling designs and examine techniques which can be used to construct them.

%%%%%%%%%%%%%%%%%%%%%%%%%%%%%%%%%%%%%%%%%%%%%%%%%%%%%%%%%%%%%%%%%%%%%%%%%%%%%%%%%%%%%%%%%
%                                                                                       %
%                                                                                       %
%%%%%%%%%%%%%%%%%%%%%%%%%%%%%%%%%%%%%%%%%%%%%%%%%%%%%%%%%%%%%%%%%%%%%%%%%%%%%%%%%%%%%%%%%
\subsection{LHD: Efficient Experimental Designs}

Latin hypercube designs (LHDs) are $n\times d$ matrices whose columns are permutations of numbers 1 to $n$ (or 0 to $n-1$) \cite{mckay1979}. They have unique point projections on every dimension and avoid replications, making them ideal for determining which inputs to use for computer experiments \cite{fang2006}. For a given number of runs and input size, an LHD can easily be generated in R:

\begin{lstlisting}
# Load an R library for finding LHDs
library(LHD)
# Generate a Latin Hypercube design with 10 runs and 2 factors
lhd1 <- rLHD(10, 2)
\end{lstlisting}

While it is intuitive to favor a design that is space-filling, in practice it is difficult to identify such designs for experiments with different number of runs and inputs. One of the more common approaches is to use orthogonal or nearly-orthogonal LHDs (OLHD). OLHDs minimize the correlations among the input settings in the design \cite{georgiou2009orthogonal} \cite{sun2017general}. They can be obtained by minimizing a correlation-based design criteria. For example, two of the most commonly used criteria for OLHDs are the average absolute correlation (ave$(|r|)$) and the maximum absolute correlation (max$|r|$):
\begin{align}
\aveF(|r|) &= \frac{2 \sum_{s=1}^{d-1} \sum_{s'=s+1}^{d}|r_{ss'}|}{d(d-1)},  \label{E4} \\
\maxF|r| &=  \maxF_{s,s'} |r_{ss'}|, \nonumber 
\end{align}
where $r_{ss'}$ is the correlation between the $s$th and $s'$th columns of the design. If the design is a true orthogonal LHD, then ave$(|r|)=0$ and max$|r|=0$. For example, to generate an OLHD in R with 8 factors and 32 runs we can write:

\begin{lstlisting}
# Obtain an orthogonal latin hypercube design 
# Need n_factor = r * 2^(C+1)
OLHD <- OLHD.S2010(C = 3, r = 2, type = "even")
\end{lstlisting}
The design can easily be verified to be orthogonal by examining:
\begin{lstlisting}
# All off diagonal elements are 0
t(OLHD) %*% OLHD
\end{lstlisting}

However, for many combinations of run size and number of inputs an orthogonal LHD does not exist, and thus a good design will be one with small ave$(|r|)$ and max$|r|$ values. Many algebraic construction methods have been proposed for finding OLHDs, and they can also be found via searching algorithms using ave$(|r|)$ or max$|r|$ as objective functions. Some specific results include \cite{ye1998}, who proposed techniques for constructing orthogonal LHDs with run-size $n=2^m$ or $n=2^m+1$ where $m$ is an integer. \cite{beattie2005} proposed to rotate the $2^d$ factorial designs for constructing $d$-factor orthogonal LHDs where $d$ must be some power of 2 and the run-size is $n=2^d$. For further examples, please refer to \cite{bursztyn2002}, \cite{steinberg2006}, \cite{cioppa2007}, \cite{lin2009}, \cite{sun2010} and \cite{yang2012}; see \cite{wang2020lhd} for a survey.
 
\begin{figure}[h]
    \centering
    \includegraphics[width=\linewidth]{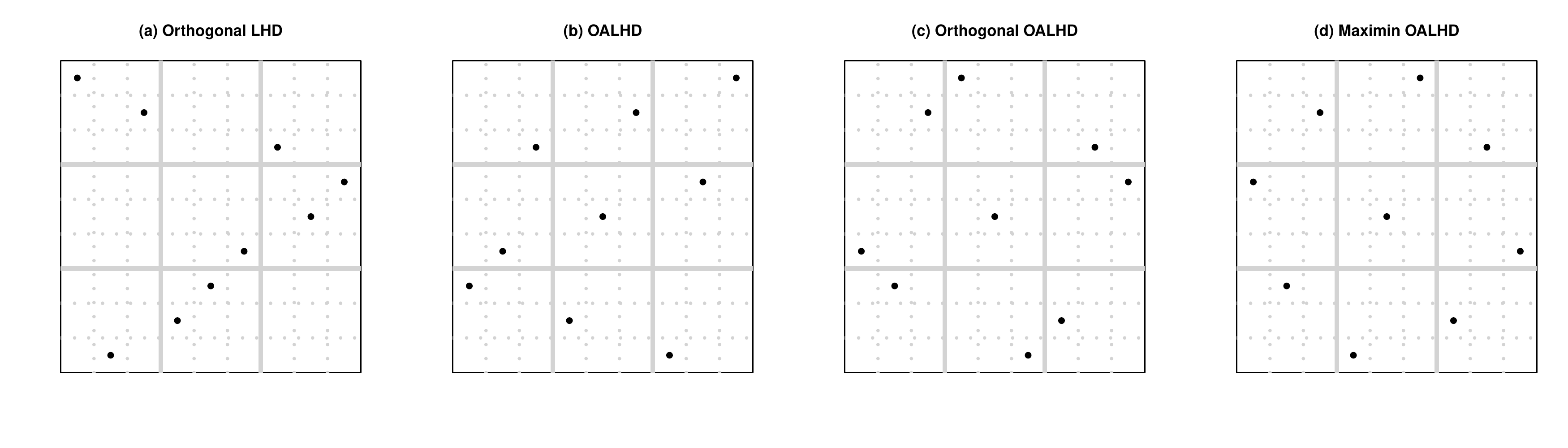}
    \caption{Some examples of 9-run 2-factor LHDs}
    \label{fig 1}
\end{figure}

While OLHDs are very commonly used, they are not guaranteed to be space-filling; see design (a) in Figure \ref{fig 1} for an example \cite{xiao2017constructions}. In light of this, various design optimality criteria have been developed related to measures of space-filling. 

\subsubsection{Centered $L_2$-Discrepancy Criteria}

\cite{Hickernell1998} defined several discrepancy based criteria among which the centered $L_2$-discrepancy (CD) is the most popular. The intuition behind the CD criteria is that a space-filling design should have points spread out uniformly in the whole design space or any sub-space of the design space. If this is the case, for any rectangular region of the design space we examine, the number of design points in that space should be proportional to the volume of that space. The CD criteria is defined as,
\begin{equation} \label{cd}
    CD(D_n)^2=\sum_{v\neq\varnothing} \int_{C^v} \left|\frac{\#(D_{n_v},J_{x_v})}{n} - \text{Volume}(J_{x_v})\right|^2\mathrm{d}x,
\end{equation}
where $D_n$ is the $n$-run, $d$-factor, $q$-level design, $v$ is some non empty subset of ${1,2,\ldots, q}$, $C^v$ is the subspace defined by the coordinate indexes selected by $v$, $D_{n_v}$ is the projection of $D_n$ onto the subspace $C^u$, $x_v$ is the projection of vector $x=(x_1,x_2,\ldots,x_q)$ on to the subspace $C^v$, $J_x$ is the chosen rectangle space defined by $x$, $J_{x_u}$ is the projection of $J_x$ onto the subspace defined by $C^u$, $\#(D_{n_u },J_{x_u})$ is the total number of designs points in $D_{n_u}$ within the chosen area defined by $J_{x_u}$, and $Volume(J_{x_u})$ is the volume of $J_{x_u}$. For more details on the rationale of the CD criteria, see the Chapter 3 in \cite{fang2006} for a survey.

\subsubsection{Multi-objective Criteria}

Another commonly-used metric for evaluating designs' space-filling properties is the maximin distance criterion \cite{Johnson1990}. This criteria favors designs with maximum pairwise distances between inputs. Maximin designs are popular due to their robustness, since the design criteria focuses on optimizing the worst case scenario $-$ the closest pairwise distance between any two points.  \cite{morris1995exploratory} defined a computationally efficient scalar value for evaluating the maximin distance criterion:

\begin{equation}
\label{phip}
\phi_p=\left(\sum_{i=2}^n \sum_{j=1}^{i-1} \frac{1}{u_{i,j}^p}\right)^\frac{1}{p} ,
\end{equation}
where $u_{i,j}$ is the distance between the $i$th and $j$th design points. Designs with smaller $\phi_p$ values are more space-filling. For sufficiently large $p$ (e.g. $p>15$), the $\phi_p$ criterion is asymptotically identical to the true maximin distance criterion.

Due to the desirability of both the orthogonality and maximin properties, \cite{joseph2008} proposed a multi-objective criterion  (denoted OMmcri) to generate orthogonal-maximin LHDs (OMm LHDs), which act as a compromise between orthogonal and maximin designs. The OMmcri criteria is given by, 

\begin{equation}
\label{OMmcri}    
\text{OMmcri}(x, \omega) =\omega \rho^2 + (1-\omega) \frac{(\phi_p-\phi_{p,lowerbound})}{(\phi_{p,upperbound}-\phi_{p,lowerbound})}.
\end{equation}
Here, $\phi_p$ is the maximin criteria value from Equation~\eqref{phip}, $\rho$ is the $ave(|r|)$ criteria value as defined in Equation~\eqref{E4}, $\omega$ is a weight value reflecting the tradeoff between the orthogonality and maximin criteria, and $\phi_{p,lowerbound}$ and $\phi_{p,upperbound}$ are given by,
\begin{align}
\phi_{p,lowerbound}&=\left\{{{n}\choose{2}}
\left(\frac{\lceil \overline{u} \rceil -  \overline{u}}{{\lfloor  \overline{u} \rfloor}^p} -
\frac{\overline{u} - \lfloor \overline{u} \rfloor}{{\lceil \overline{u} \rceil}^p}\right)\right\}^{\frac{1}{p}}, \text{ and} \nonumber \\
\phi_{p,upperbound} &= \left(\sum_{i=1}^{n-1} \frac{n-i}{(id)^p}\right)^{\frac{1}{p}}, \nonumber
\end{align}
respectively. Here $\overline{u}$ is the average distance between the design points and $ \lfloor \overline{u} \rfloor$ and $\lceil \overline{u} \rceil$ are the largest integer smaller than $\overline{u}$ and the smallest integer  larger than $ \overline{u}$.

Another popular class of efficient LHDs is the orthogonal array based LHDs (OALHDs) by \cite{tang1993}, where the levels in randomized orthogonal arrays (OAs) are expanded to form LHDs. The OALHDs have desirable sampling and projection properties, but they are not necessarily space-filling \cite{xiao2017constructions}; see designs (b) and (c) in Figure \ref{fig 1} for some examples. \cite{leary2003} proposed to use a simulated annealing algorithm to search for space-filling OALHDs, and \cite{xiao2018} further proposed to consider both level permutation and level expansion for generating OALHDs. Some algebraic construction methods are also available for constructing maximin LHDs for certain design sizes \cite{xiao2017} \cite{wang2018}.

\subsubsection{Maxpro: Maximum projection designs}

Space-filling LHDs, including CD and and maximin distance LHDs, focus on the design's properties in the full dimensional spaces. Yet, their space-filling properties in some sub-spaces (projections) may not be adequate.  \cite{joseph2015maximum} proposed the maximum projection LHDs (Maxpro LHDs) that guarantee designs have space-filling properties in all projections. The maximum projection criterion is defined as  
	\begin{equation}\label{E3}
	\minF_{\textbf{X}} \psi (\textbf{X}) = \Bigg\{ \frac{1}{{n \choose 2}} \sum_{i=1}^{n-1} \sum_{j=i+1}^{n} \frac{1}{\prod_{s=1}^{d}(x_{is}-x_{js})^2}  \Bigg\}^{1/d}.
	\end{equation}
Here, $\textbf{X}$ is a $n \times d$ matrix where each row is an input to the computer experiment, and the minimization is over all pairs of rows in $\textbf{X}$. Clearly a design minimizing $\psi$ will have every pair of design points  apart from each other in all projections, justifying the name ``Maxpro."

%%%%%%%%%%%%%%%%%%%%%%%%%%%%%%%%%%%%%%%%%%%%%%%%%%%%%%%%%%%%%%%%%%%%%%%%%%%%%%%%%%%%%%%%%
%                                                                                       %
%                                                                                       %
%%%%%%%%%%%%%%%%%%%%%%%%%%%%%%%%%%%%%%%%%%%%%%%%%%%%%%%%%%%%%%%%%%%%%%%%%%%%%%%%%%%%%%%%%
\subsection{Searching Algorithms for Generating Efficient LHDs with Flexible Sizes}
\label{Algs}

The metrics discussed above for evaluating designs such as the minimax criteria provide a way of quantifying how ``good" a design is in some sense. It remains to determine how to actually construct designs that have a good value of the criterion, which is a challenging problem in many situations. For many such design problems, it is popular to use metaheuristic optimization algorithms to find designs. Metaheuristic algorithms are often used to solve problems in astronomy, see, for example, \cite{charbonneau1995genetic}, \cite{giuliano2008multi}, \cite{mohanty2012particle}, \cite{misiak2016evolutionary}, and \cite{mohanty2020adaptive}. They can be applied to solve difficult problems such as clustering in complex data \cite{djorgovski2003challenges}, \cite{hruschka2009survey}.

These algorithms are preferred due to their flexibility - in general they will work with any objective function. For a more detailed review of metaheuristic algorithms for finding designs, see \cite{mandal2015algorithmic}. Here we will focus on two of the more commonly used approaches: Simulated Annealing and the Genetic Algorithm.

\subsubsection{Simulated Annealing Algorithms} 
	
Simulated Annealing (SA) is one of the most widely used general probabilistic optimization techniques \cite{kirkpatrick1983optimization}. The algorithm follows the annealing process in metallurgy, in which materials are heated to a high temperature where their properties change, and then are allowed to slowly cool. \cite{morris1995exploratory} adapted the classic SA algorithm for finding maximin distance LHDs, and the approach can easily be modified to search for other types of designs by using the other optimality criteria defined in Section 3.1. 
	
SA starts with a random LHD and then improves it via an element exchange method, where  two random elements from a random column in the design are exchanged. If this exchange  results in a more efficient design, then the change is kept. If the exchange does not result in any improvement, the change is kept with probability controlled by the current temperature (tuning parameter). Allowing changes that do not improve the design helps the search algorithm to escape local optima. The SA algorithm will iteratively repeat this exchange procedure. After a certain number of rounds, the temperature would be annealed to decrease (cool down) the probability of updating the current design following the annealing schedule. We summarize a general SA framework in Algorithm~\ref{Alg1}, where the target function $\Phi$ to be minimized can be the optimality criterion defined in \eqref{E4}, \eqref{cd}, \eqref{phip}, \eqref{OMmcri} and \eqref{E3} for the orthogonal, CD, maximin, OMm and Maxpro LHDs, respectively.

In the Algorithm~\ref{Alg1}, the maximum number of iterations $N$ is recommended to be around 500 according to the convergence analysis in \cite{wang2020lhd}. The decreasing rate for the current temperature $T$ is another important tuning parameter. A larger rate will make $T$ decline faster, and thus lead to a faster stop of the algorithm. Yet, it may also result in larger probability of missing the true global optimum. Considering this trade-off, it is recommended to set $T$ between $0.05$ to $0.15$. The tuning parameter $S$  indicates the maximum consecutive attempts the algorithm will try without improvements before temperature reduces, and \cite{morris1995exploratory} recommends it to be around 5, depending on how expensive the objective function is to evaluate.
	
It is straightforward to use Simulated Annealing to find designs in R. For example:
\begin{lstlisting}
# 10 Runs, 2 inputs, 25 iterations of Simulated Annealing algorithm
LHD_SA <- SA(n = 10, k = 2, N = 25)
\end{lstlisting}
Similarly, designs satisfying the multi-objective approach can be found by:
\begin{lstlisting}
# 10 Runs, 2 inputs, 25 iterations of Simulated Annealing algorithm
# using multi-objective
multi_obj_design <- SA2008(n = 10, k = 2, N = 25)
\end{lstlisting}

\subsubsection{Genetic Algorithms}
	
The genetic algorithm (GA) is a metaheuristic algorithm inspired by the process of natural selection \cite{holland1992adaptation} \cite{goldberg1989genetic}. The GA starts from a population of randomly generated candidate solutions (designs), called chromosomes. The population of chromosomes in each iteration is called a generation. For each generation the objective function will be evaluated for each chromosome, with the corresponding value being known as the fitness. The more fit chromosomes will be allowed to survive to the next generation, while the less fit chromosomes will be replaced by new offspring. These offspring are obtained by selecting several chromosomes (called parents) and recombining their settings using crossover and mutation techniques to produce offspring with potentially better fitness.

\cite{liefvendahl2006study} adapted the general GA framework for searching for maximin LHDs. Their approach begins with random LHDs as the initial population. They then perform a selection step in which the best half of the LHDs are allowed to survive to the next generation. Then, a crossover step is performed in which random columns in these survivors are exchanged with other survivors. Additionally, to encourage diversity in the solutions and prevent the algorithm becoming stuck in a local optima, a mutation step is performed in which two random elements in a column are exchanged. Note that the current best chromosome is excluded from this mutation in order to preserve the best current solution. Finally, the fitness of the new population of LHDs is calculated, and the process is repeated until the stopping criteria is satisfied. We include a detailed description of the GA, along with the tuning parameters, in Algorithm~\ref{Alg3}.

\begin{algorithm}
		\caption{Simulated Annealing for LHD}
		\begin{algorithmic}[1]\label{Alg1}
			\STATE Choose values for the tuning parameters: the starting temperature $T$, the number of attempts before lowering the temperature $S$, and the maximum number of iterations $N$.
			\STATE Set the counter index $C=1$.
			\STATE Construct a random starting LHD $\textbf{X}$.
			\STATE Select a column from $\textbf{X}$ at random.
			\STATE Exchange two randomly selected elements within this chosen column. Denote the new design by $\textbf{X}_{new}$.
			\STATE If $\Phi(\textbf{X}_{new}) < \Phi(\textbf{X})$, then $\textbf{X}=\textbf{X}_{new}$ (accept the new design). Otherwise, let $\textbf{X}=\textbf{X}_{new}$ with probability  $\hbox{exp}\left\{-\frac{\Phi(\textbf{X}_{new})-\Phi(\textbf{X})}{T}\right\}$.
			\STATE If $S$ attempts have passed since the last improvement, decrease the temperature $T$ and repeat Steps 4$-$6.
			\STATE If $C < N$, increment
			$C$ and repeat Steps 4$-$7; Otherwise, terminate and return $\textbf{X}$.
		\end{algorithmic}
\end{algorithm}
	
\begin{algorithm}
		\caption{Genetic Algorithm for LHD}
		\begin{algorithmic}[1]\label{Alg3}
			\STATE Set the probability of mutation, $p_{mut}$. Suggested setting is $1 / (d-1)$ \cite{liefvendahl2006study}. Set the maximum number of iterations $N$ and the counter index $C=1$.
			\STATE Generate $m$ random $n \times d$ LHDs, denoted by $X_{1},\ldots, X_{m}$, where $m$ is the population size (number of chromosomes). Here, $m$ must be an even number.
			\STATE Evaluate the objective function, $\Phi(X_{i})$, for $i=1,\ldots, m$.
			
			\STATE Select ${\it survivors}$: order the $X_i$ by their objective function values and select the best $\frac{m}{2}$ $X_{i}$ (with the smallest $\frac{m}{2}$ $\Phi$ values), denoted by $X_{i}^{s}$ for $i=1,\ldots, \frac{m}{2}$, WLOG.
			\STATE Let $X_{b}^{s} =\argmin_{i} \Phi(X_{i}^{s})$ (i.e. $X_{b}^{s}$ is the best ${\it survivor}$)
			
			\FOR{each $X_{i}^{s}$, excluding $X_{b}^{s}$,}
			\STATE Randomly choose a column $j$ from $X_{b}^{s}$, and replace it with the $j^{th}$ column from $X_{i}^{s}$. 
			\ENDFOR
			
			\FOR{each $X_{i}^{s}$, excluding $X_{b}^{s}$,}
			\STATE Randomly choose a column $j$ from $X_{i}^{s}$, and replace it with the $j^{th}$ column from $X_{b}^{s}$.
			\ENDFOR
			
			\STATE Update $X_{i}$: let $X_{1}=X_{b}^{s}$ and the  $X_{2},\ldots,X_{m/2}$ be the design matrices obtained by steps 6$-$8. Let $X_{m/2+1}=X_{b}^{s}$ and $X_{m/2+2},\ldots,X_{m}$ be the design matrices generated by Steps 9$-$11. 
			
			\FOR{each $X_{i}$ (except $X_{1}$)}
			\FOR{each column $j$ of $X_{i}$}
			\IF{$z < p_{mut}$ where $z \sim \text{Uniform}(0,1)$ }
			\STATE Exchange two randomly selected elements in $j$.
			\ENDIF
			\ENDFOR
			\ENDFOR
			
			\STATE Calculate $\Phi(X_{i})$ for all $i$.
			
			\STATE if $C \leq N$, set $C=C+1$ and repeat Steps 4-21;  otherwise, stop the algorithm.
		\end{algorithmic}
\end{algorithm} 

It is also straightforward to use the GA to find space-filling designs in R. For example:
\begin{lstlisting}
# 10 Runs, 2 inputs, 25 iterations of Genetic algorithm
# OC is optimality criteria - phi_p is the maximin distance
LHD_GA <- GA(n = 10, k = 2, N = 25, OC = "phi_p")
\end{lstlisting}

%%%%%%%%%%%%%%%%%%%%%%%%%%%%%%%%%%%%%%%%%%%%%%%%%%%%%%%%%%%%%%%%%%%%%%%%%%%%%%%%%%%%%%%%%
%                                                                                       %
%                                                                                       %
%                                                                                       %
%                                     SECTION 4                                         %
%                                                                                       %
%                                                                                       %
%                                                                                       %
%%%%%%%%%%%%%%%%%%%%%%%%%%%%%%%%%%%%%%%%%%%%%%%%%%%%%%%%%%%%%%%%%%%%%%%%%%%%%%%%%%%%%%%%%
\section{Summary and Conclusions}

Sophisticated computer simulators allow scientists to test complex systems which would be too expensive or completely impossible to assess otherwise. These simulations are usually very time-consuming, and computationally cheap surrogates are called for to facilitate the analysis and optimization of the underlying system. Gaussian processes are popular choices for such surrogates (or emulators). In order to effectively reap the benefits of utilizing the surrogate, the simulator should be evaluated on a set of points chosen efficiently. Latin hypercube designs have proven efficient for that purpose. 

In this tutorial paper we discussed design criteria and subsequent metaheuristic optimization strategies for finding designs that allow astronomy researchers to extract the maximum benefit offered by Gaussian process surrogate modeling. We provided an overview of model fitting using Gaussian processes and identification of optimal Latin hypercube designs. Relevant R codes have been used for illustration. Apart from the libraries discussed in the paper, there are many other packages in R that can be used. Interested readers may want to consider the laGP (Local Approximate Gaussian Process Regression \cite{r:gramacy2019}), DiceKriging (Kriging Methods for Computer Experiments \cite{r:roustant2020}), GPfit (Gaussian Processes Modeling \cite{r:mac2019}) and SLHD (Maximin-Distance (Sliced) Latin Hypercube Designs \cite{r:ba2015}) packages.

One consideration not covered in this tutorial paper is how to best utilize Gaussian process models when the data sets are astronomically large. Such ``big data" may cause the estimation techniques to become quite slow, requiring advanced techniques to speed up the estimation. This is a topic of active research. For further details, see \cite{liu2020gaussian}.


\begin{thebibliography}{}

\bibitem{r:ba2015}
Ba, S. (2015), SLHD: Maximin-Distance (Sliced) Latin Hypercube Designs, URL:\url{https://cran.r-project.org/web/packages/SLHD/index.html}, R package version 2.1-1.

\bibitem{Ba2012}
Ba, S. and  Joseph, V. R. (2012),  Composite Gaussian process models for emulating expensive functions, \emph{   Ann. Appl. Stat.}, {\bf 6}, 4, 1838--1860.

\bibitem{beattie2005}
  {Beattie, S. D. and Lin, D. K. J.},
   {(2005)},
  {A new class of Latin hypercube for computer experiments},
  \emph{Contemporary Multivariate Analysis and Designs of Experiments, in Celebration of Prof. Kai-Tai Fang’s 65th Birthday. Singapore: World Scientific},
   {205--226}.
   
 \bibitem{behroozi2019universemachine}
 {Behroozi, P., Wechsler, R., Hearin, A., and Conroy, C.},
 {(2019)},
 {UniverseMachine: The correlation between galaxy growth and dark matter halo assembly from z= 0$-$10},
 \emph{Monthly Notices of the Royal Astronomical Society},
 {\bf 488}, 3, 3143--3194, Oxford University Press.



\bibitem{natalia}
 Bhattacharjeea, N., Ranjan, P., Mandal, A. and Tollner, E. W. (2019), A history matching approach for calibrating hydrological models,  \emph{Environmental and Ecological Statistics}, {\bf 26}, 1, 87--105.

\bibitem{Derek2}
Bingham, D., Branin Function. \emph{Virtual Library of Simulation Experiments}, \url{https://www.sfu.ca/~ssurjano/branin.html}.

\bibitem{bursztyn2002}
  {Bursztyn, D. and Steinberg, D. M.},
   {(2002)},
  {Rotation designs: orthogonal first-order designs with higher order projectivity},
  \emph{Applied Stochastic Models in Business and Industry},
  {\bf 18},
  {3},
   {197--206},
   {Wiley Online Library}.


% % REMOVE REFERENCE
% \bibitem{chen2013optimizing}
% 	{Chen, R.-B., Hsieh, D.-N. and Hung, Y. and Wang, W.},
% 	 {(2013)},
% 	{Optimizing {L}atin hypercube designs by particle swarm},
% 	\emph{Statistics and computing},
% 	{\bf 23},
% 	{5},
% 	 {663--676},
% 	 {Springer}.
\bibitem{charbonneau1995genetic}
    {Charbonneau, P.},
    {(1995)},
    {Genetic algorithms in astronomy and astrophysics},
    \emph{The Astrophysical Journal Supplement Series},
    {\bf 101},
    {309--334}.

\bibitem{cioppa2007}
  {Cioppa, T. M. and Lucas, T. W.},
   {(2007)},
  {Efficient nearly orthogonal and space-filling Latin hypercubes},
  \emph{Technometrics},
  {\bf 49},
  {1},
   {45--55},
   {Taylor \& Francis}.


\bibitem{r:dancik2020}
Dancik, G. M. (2020), mlegp: Maximum Likelihood Estimates of Gaussian Processes, URL:\url{https://cran.r-project.org/web/packages/mlegp/index.html}, R package version 3.1.8.


\bibitem{deng2017additive}
	{Deng, X., Lin, C. D., Liu, K.-W. and Rowe, R. K.},
	 {(2017)},
	{Additive Gaussian process for computer models with qualitative and quantitative factors},
	\emph{Technometrics},
	{\bf 59},
	{3},
	 {283--292},
	 {Taylor \& Francis}.
	 
\bibitem{djorgovski2003challenges}
    {Djorgovski, SG., Brunner, R., Mahabal, A., Williams, R., Granat, R., and Stolorz, P.},
    (2003),
    Challenges for cluster analysis in a virtual observatory,
    {Statistical Challenges in Astronomy},
    {127--141},
    {Springer}.


% % REMOVE REFERENCE
% \bibitem{Fang2005}
% Fang, K.-T., Li, R. and Sudjianto, A. (2005),  Design and modeling for computer experiments. CRC press.

\bibitem{fang2006}
  {Fang, K. T., Li, R. and Sudjianto, A.},
   {(2006)},
  {Design and modeling for computer experiments},
   {CRC Press}.
   
\bibitem{forrester2008engineering}
  {Forrester, A., Sobester, A., and Keane, A.},
  (2008)
  {Engineering design via surrogate modelling: a practical guide},
  {John Wiley \& Sons}.


\bibitem{georgiou2009orthogonal}
	{Georgiou, S. D.},
	 {(2009)},
	{Orthogonal {L}atin hypercube designs from generalized orthogonal designs},
	\emph{Journal of Statistical Planning and Inference},
	{\bf 139},
	{4},
	 {1530--1540},
	 {Elsevier}.

\bibitem{giuliano2008multi}	 
    {Giuliano, M., and Johnston, M.},
    {(2008)},
    \emph{Multi-Objective Evolutionary Algorithms for Scheduling the James Webb Space Telescope.},
    {ICAPS},
    {107--115}.


% this is right...
\bibitem{goldberg1989genetic}
	{Goldberg, D. E.},
	 {(1989)},
	{Genetic algorithms in search},
	\emph{Optimization and MachineLearning},
	 {Addison Wesley Publishing Co. Inc.}


\bibitem{Gramacy2020}
Gramacy, R. B. (2020). Surrogates: Gaussian Process Modeling, Design, and Optimization for the Applied Sciences. CRC Press.


\bibitem{Gramacy2015}
Gramacy, R. B. and Apley, D. W. (2015),  Local Gaussian Process Approximation for Large Computer Experiments, \emph{Journal of Computational and Graphical Statistics}, {\bf 24}, 2.


\bibitem{Gramacy2008}
Gramacy, R. B. and Lee, H. K. H. (2008),  Bayesian treed Gaussian process models with an application to computer modeling. \emph{J. Amer. Statist. Assoc.}, {\bf 103}, 1119–-1130.

\bibitem{r:gramacy2019}
Gramacy, R. B., Sun, F. (2019), laGP: Local Approximate Gaussian Process Regression, URL:\url{https://cran.r-project.org/web/packages/laGP/index.html}, R package version 1.5-5.

\bibitem{gration2019dynamical}
{Gration, A. and Wilkinson, M.},
{(2019)},
{Dynamical modelling of dwarf spheroidal galaxies using Gaussian-process emulation},
\emph{Monthly Notices of the Royal Astronomical Society},
{\bf 485}, 4, {4878--4892}, {Oxford University Press}.

\bibitem{han2009}
	{Han, G., Santner, T. J., Notz, W. I., Bartel, D. L.},
	 {(2009)},
	{Prediction for computer experiments having quantitative and qualitative input variables},
	\emph{Technometrics},
	{\bf 51},
	{3},
	 {278--288},
	 {Taylor \& Francis}.


\bibitem{Hickernell1998}
  {Hickernell, F.},
   {(1998)}
  {A generalized discrepancy and quadrature error bound},
  \emph{Mathematics of Computation of the American Mathematical Society},
  {\bf 67},
  {221},
   {299--322}.


\bibitem{holland1992adaptation}
	{Holland, J. H. and others},
	 {(1992)},
	{Adaptation in natural and artificial systems: an introductory analysis with applications to biology, control, and artificial intelligence},
	 {MIT press}.

\bibitem{hruschka2009survey}
    {Hruschka, E., Campello, R., Freitas, A., and others},
    (2009)
    {A survey of evolutionary algorithms for clustering},
    \emph{IEEE Transactions on Systems, Man, and Cybernetics, Part C (Applications and Reviews)},
    {\bf 39},
    {2},
    {133--155},
    {IEEE}


\bibitem{Johnson1990}
  {Johnson, M. E., Moore, L. M. and Ylvisaker, D.},
   {(1990)},
  {Minimax and maximin distance designs},
  \emph{Journal of Statistical Planning and Inference},
  {\bf 26},
  {2},
   {131--148},
   {Elsevier}.


\bibitem{joseph2008}
  {Joseph, V. R. and Hung, Y.},
   {(2008)},
  {Orthogonal-maximin Latin hypercube designs},
  \emph{Statistica Sinica},
   {171--186},
   {JSTOR}.

% % remove reference
% \bibitem{Roshan2008}
% Joseph, V. R., Hung, Y., and Sudjianto, A. (2008),  Blind Kriging: A New Method for Developing Metamodels. \emph{ASME Journal of Mechanical Design}, {\bf 130}, 031102-1-8.


\bibitem{joseph2015maximum}
  {Joseph, V. R., Gul, E. and Ba, S.},
   {(2015)},
  {Maximum projection designs for computer experiments},
  \emph{Biometrika},
  {\bf 102},
  {2},
   {371--380},
   {Oxford University Press}.



\bibitem{KOH}
Kennedy, M. and O'Hagan, A. (2002), Bayesian calibration of computer models, \emph{Journal of the Royal Statistical Society Series B (Statistical Methodology)}. {\bf 63}, 3, 425--464, Wiley Online Library.

% \bibitem{kidder2000}
% Kidder, L.E., Scheel, M.A., Teukolsky, S.A., Carlson, E.D. and Cook, G.B., (2000), Black hole evolution by spectral methods. Physical Review D, {\bf 62}, 8, 084032. Vancouver	.


\bibitem{kirkpatrick1983optimization}
	{Kirkpatrick, S., Gelatt, C. D. and Vecchi, M. P.},
	 {(1983)},
	{Optimization by simulated annealing},
	\emph{Science},
	{\bf 220},
	{4598},
	 {671--680},
	 {American Association for the Advancement of Science}.

\bibitem{kruckow2018progenitors}
{Kruckow, M., Tauris, T., Langer, N., Kramer, M., and Izzard, Robert G},
  {(2018)},
  {Progenitors of gravitational wave mergers: binary evolution with the stellar grid-based code COMBINE},
  \emph{Monthly Notices of the Royal Astronomical Society},
  {\bf 481},2, {1908--1949}, {Oxford University Press}.


\bibitem{leary2003}
  {Leary, S., Bhaskar, A., and Keane, A.},
   {(2003)},
  {Optimal orthogonal-array-based latin hypercubes},
  \emph{Journal of Applied Statistics},
  {\bf 30},
  {5},
   {585--598},
   {Taylor \& Francis}.


\bibitem{liefvendahl2006study}
	{Liefvendahl, M. and Stocki, R.},
	 {(2006)},
	{A study on algorithms for optimization of {L}atin hypercubes},
	\emph{Journal of Statistical Planning and Inference},
	{\bf 136},
	{9},
	 {3231--3247},
	 {Elsevier}.


% % REMOVE REFERENCE
% \bibitem{Derek1}
% Lin, L., Barrett, J., Bingham, D., and Mandel, I. Uncertainty Quantification of Computer Model for Binary Black Hole Formation. \url{http://www.sfu.ca/~lla24/UQ_BBH_Luyao.pdf}.



\bibitem{lin2009}
  {Lin, C. D., Mukerjee, R. and Tang, B.},
   {(2009)},
  {Construction of orthogonal and nearly orthogonal Latin hypercubes},
  \emph{Biometrika},
   {243--247},
   {JSTOR}.



\bibitem{liu2020gaussian}
{Liu, H., Ong, Y-S., Shen, X. and Cai, J.},
{(2020)},
{When Gaussian process meets big data: A review of scalable GPs},
\emph{IEEE transactions on neural networks and learning systems},
{\bf 31},
	{11},
	 {4405--4423},
	 {IEEE}.


\bibitem{mandal2015algorithmic}
Mandal, A., Wong, W. K., and Yu, Y. (2015), Algorithmic searches for optimal designs, \emph{Handbook of design and analysis of experiments},{755--783}, {CRC Press Boca Raton, FL}.


\bibitem{r:mac2019}
MacDoanld, B., Chipman, H., Campbell, C. and Ranjan, P. (2019), GPfit: Gaussian Processes Modeling, URL:\url{https://cran.r-project.org/web/packages/GPfit/index.html}, R package version 1.0-8.



\bibitem{mckay1979}
  {McKay, M. D. and Beckman, R. J. and Conover, W. J.},
   {(1979)},
  {Comparison of three methods for selecting values of input variables in the analysis of output from a computer code},
  \emph{Technometrics},
  {\bf 21},
  {2},
   {239--245},
   {Taylor \& Francis}.
   
\bibitem{misiak2016evolutionary}
  {Misiak, M., and others},
  {(2016)}
  {Evolutionary Algorithms in Astrodynamics},
  \emph{International Journal of Astronomy and Astrophysics},
  {\bf 6},
  {04},
  {435--439},
  {Scientific Research Publishing}.


% % remove reference
% \bibitem{Morris1995}
%   {Morris, M. D. and Mitchell, T. J.},
%   {(1995)},
%   {Exploratory designs for computational experiments},
%   \emph{Journal of Statistical Planning and Inference},
%   {\bf 43},
%   {3},
%   {381--402},
%   {Elsevier}.

\bibitem{mohanty2012particle}
    {Mohanty, S.},
    {(2012)}
    {Particle Swarm Optimization and regression analysis--I},
    \emph{Astronomical Review},
    {\bf 7},
    {2},
    {29--35},
    {Taylor \& Francis}

\bibitem{mohanty2020adaptive}
    {Mohanty, S. and Fahnestock, E.},
    {(2020)},
    {Adaptive spline fitting with particle swarm optimization},
    \emph{Computational Statistics},
    {\bf 36},
    pages={155--191},
    {Springer}.

\bibitem{morris1995exploratory}
	{Morris, M. D. and Mitchell, T. J.},
	 {(1995)},
	{Exploratory designs for computational experiments},
	\emph{Journal of statistical planning and inference},
	{\bf 43},
	{3},
	 {381--402},
	 {Elsevier}.


% % Remove Reference
% \bibitem{Parussini2017}
% Parussini, L., Venturi, D., Perdikaris, P. and Karniadakis, G.E., 2017. Multi-fidelity Gaussian process regression for prediction of random fields. {\it Journal of Computational Physics}, {\bf 336}, 36--50.



\bibitem{qian2008}
	{Qian, P. Z. G. and Wu, H. and Wu, C. F. J.},
	 {(2008)},
	{Gaussian process models for computer experiments with qualitative and quantitative factors},
	\emph{Technometrics},
	{\bf 50},
	{3},
	 {383--396},
	 {Taylor \& Francis}.
	 
\bibitem{RSoftware}
{{R Core Team}},
{R: A Language and Environment for Statistical Computing},
{R Foundation for Statistical Computing},
{Vienna, Austria},
{2019},
{https://www.R-project.org/}.


\bibitem{joseph}
Resch, J.,  Mandal, A. and Ranjan, P. (2021), ``Inverse problem for dynamic computer simulators via multiple scalar-valued contour estimation'', \url{https://arxiv.org/abs/2010.08941}.

\bibitem{r:roustant2020}
Roustant, O., Ginsbourger, D., Deville, Y., Clement, C. and Richet, Y. (2020), DiceKriging: Kriging Methods for Computer Experiments, URL:\url{https://cran.r-project.org/web/packages/DiceKriging/index.html}, R package version 1.5.8.


\bibitem{Sacks89}
Sacks, J., Welch, W. J., Mitchell, T. J., and Wynn, H. P. (1989),  Design and analysis of computer experiments. \emph{Statistical science}, 409--423.

% % Remove reference
% \bibitem{Singh2011}
% Singh, R. K., Joseph, V. R. and Melkote, S. N., (2011),  A statistical approach to the optimization  of  a  laser-assisted  micromachining  process. \emph{The  International  Journal  of  Advanced Manufacturing Technology}, {\bf 53}, 221--230.



\bibitem{steinberg2006}
  {Steinberg, D. M. and Lin, D. K. J.},
   {(2006)},
  {A construction method for orthogonal Latin hypercube designs},
  \emph{Biometrika}, {\bf 93}, 2, 
   {279--288},
   {JSTOR}.


\bibitem{Stevenson1}
Stevenson, S., Vigna-Gómez, A., Mandel, I., Barrett, J.W., Neijssel, C.J., Perkins, D. and De Mink, S.E., (2017),  Formation of the first three gravitational-wave observations through isolated binary evolution. Nature Communications, {\bf 8}, 1, 1--7. Vancouver.	


\bibitem{sun2010}
  {Sun, F., Liu, M.-Q. and Lin, D. K. J.},
   {(2010)},
  {Construction of orthogonal Latin hypercube designs with flexible run sizes},
  \emph{Journal of Statistical Planning and Inference},
  {\bf 140},
  {11},
   {3236--3242},
   {Elsevier}.


\bibitem{sun2017general}
	{Sun, F. and Tang, B.},
	 {(2017)},
	{A general rotation method for orthogonal {L}atin hypercubes},
	\emph{Biometrika},
	{\bf 104},
	{2},
	 {465--472},
	 {Oxford University Press}.


% % Remove reference
% \bibitem{Sung2020}
% Sung, C.L., Hung, Y., Rittase, W., Zhu, C. and Jeff Wu, C.F., (2020),  A generalized Gaussian process model for computer experiments with binary time series. {\it Journal of the American Statistical Association}, {\bf 115}, 530, 945--956, Vancouver.




\bibitem{swiler2014surrogate}
	{Swiler, L. P. and Hough, P. D. and Qian, P. Z. G., Xu, X., Storlie, C. and Lee, H.},
	 {(2014)},
	{Surrogate models for mixed discrete-continuous variables},
	{Constraint Programming and Decision Making},
	 {181--202},
	 {Springer}.



\bibitem{tang1993}
  {Tang, B.},
   {(1993)},
  {Orthogonal array-based Latin hypercubes},
  \emph{J. Amer. Statist. Assoc.},
  {\bf 88},
  {424},
   {1392--1397},
   {Taylor \& Francis}.


\bibitem{vernon2010}
	{Vernon, I., Goldstein, M. and Bower, R. G.},
	 {(2010)},
	{Galaxy formation: a Bayesian uncertainty analysis},
	\emph{Bayesian Analysis},
	{\bf 5},
	{4},
	 {619--669},
	 {International Society for Bayesian Analysis}.


\bibitem{COMPAS2}
Vigna-Gómez, A., Neijssel, C.J., Stevenson, S., Barrett, J.W., Belczynski, K., Justham, S., de Mink, S.E., Müller, B., Podsiadlowski, P., Renzo, M. and Szécsi, D., 2018. On the formation history of Galactic double neutron stars. \emph{Monthly Notices of the Royal Astronomical Society}, {\bf 481}, 3, 4009--4029. Vancouver.


\bibitem{Wackernagel}
Wackernagel H. (2002), Multivariate Geostatistics, Springer.


\bibitem{r:wang2020}
Wang, H., Xiao, Q., and Mandal, A., (2020), LHD: Latin Hypercube Designs (LHDs), URL:\url{https://CRAN.R-project.org/package=LHD}, R package version 1.3.1.



\bibitem{wang2020lhd}
  {Wang, H., Xiao, Q. and Mandal, A.},
   {(2020)}
  {Musings about Constructions of Efficient Latin Hypercube Designs with Flexible Run-sizes},
  \emph{arXiv preprint arXiv:2010.09154v2}.



\bibitem{wang2018}
  {Wang, L., Xiao, Q. and Xu, H.},
   {(2018)},
  {Optimal maximin $L_1$-distance Latin hypercube designs based on good lattice point designs},
  \emph{Annals of Statistics},
  {\bf 46},
  {6B},
   {3741--3766},
   {Institute of Mathematical Statistics}.




\bibitem{Welch1992}
Welch, W. J., Buck, R. J., Sacks, J., Wynn, H. P., Mitchell, T. J. and Morris, M. D. (1992),  Screening, predicting, and computer experiments, \emph{Technometrics}, {\bf 34}, 15–-25.


% % Remove reference
\bibitem{williams2019}
Williams, D., Heng, I.S., Gair, J., Clark, J.A. and Khamesra, B., (2019),  A Precessing Numerical Relativity Waveform Surrogate Model for Binary Black Holes: A Gaussian Process Regression Approach. arXiv preprint arXiv:1903.09204.




\bibitem{xiao2017constructions}
  {Xiao, Q.},
   {(2017)},
  {Constructions and Applications of Space-Filling Designs},
  {Ph.D. Dissertation, University of California Los Angeles}.




\bibitem{Qian1}
Xiao, Q., Mandal, A., Lin, C. D., and Deng, X. (2021),  EzGP: Easy-to-interpret Gaussian Process models for computer experiments with both quantitative and qualitative factors. under revision for  \emph{SIAM/ASA Journal on Uncertainty Quantification}.



% % Remove refeerence
% \bibitem{Qian2}
% Xiao, Q.; Wang, Y.; Mandal, A. and Deng, X., (2021),  Modelling and active learning for experiments with quantitative-sequence factors. under revision for  \emph{Journal of American Statistical Association $-$ Theory and Methods}.


\bibitem{xiao2017}
  {Xiao, Q. and Xu, H.},
   {(2017)},
  {Construction of maximin distance Latin squares and related Latin hypercube designs},
  \emph{Biometrika},
  {\bf 104},
  {2},
   {455--464},
   {Oxford University Press}.


\bibitem{xiao2018}
  {Xiao, Q. and Xu, H.},
   {(2018)},
  {Construction of maximin distance designs via level permutation and expansion},
  \emph{Statistica Sinica},
  {\bf 28},
  {3},
   {1395--1414},
   {JSTOR}.


\bibitem{xiong2007}
Xiong, Y., Chen, W., Apley, D. W. and Ding, X. (2007),  A non-stationary covariance-based kriging method for metamodelling in engineering design. \emph{Internat. J. Numer. Methods Engrg.}, {\bf 71}, 733-–756.




\bibitem{yang2012}
  {Yang, J. and Liu, M.-Q.},
   {(2012)},
  {Construction of orthogonal and nearly orthogonal Latin hypercube designs from orthogonal designs},
  \emph{Statistica Sinica},
   {433--442},
   {JSTOR}.



\bibitem{ye1998}
  {Ye, K. Q.},
   {(1998)},
  {Orthogonal column Latin hypercubes and their application in computer experiments},
  \emph{Journal of the American Statistical Association},
  {\bf 93},
  {444},
   {1430--1439},
   {Taylor \& Francis}.



\bibitem{zhang2015computer}
	{Zhang, Y. and Notz, W. I.},
	 {(2015)},
	{Computer experiments with qualitative and quantitative variables: a review and reexamination},
	\emph{Quality Engineering},
	{\bf 27},
	{1},
	 {2--13},
	 {Taylor \& Francis}.



\bibitem{zhang2018latent}
	{Zhang, Y., Tao, S., Chen, W. and Apley, D. W.},
	{(2019)},
	{A latent variable approach to Gaussian process modeling with qualitative and quantitative factors},
	\emph{Technometrics},
	 {1--12},
	 {Taylor \& Francis}.



\bibitem{zhou2011}
	{Zhou, Q. and Qian, P. Z. G. and Zhou, S.},
	 {(2011)},
	{A simple approach to emulation for computer models with qualitative and quantitative factors},
	\emph{Technometrics},
	{\bf 53},
	{3},
	 {266--273},
	 {Taylor \& Francis}.


\end{thebibliography}
\end{document}